\documentclass[a4paper]{paper}
\usepackage{graphicx}

\begin{document}
\title{Integrability of Three Dimensional Gravity Field Equations}

\small{
\author{Metin G{\" u}rses\\
Department of Mathematics, Faculty of Sciences, Bilkent University, \\
06800 Ankara, Turkey}}


\maketitle

\begin{abstract}
We show that the tree dimensional Einstein vacuum field equations with cosmological constant are integrable. Using the $sl(2,R)$ valued soliton connections we obtain the metric of the spacetime in terms of the dynamical variables of the integrable nonlinear partial differential equations.
\end{abstract}


\section{Introduction}
Gravity in three dimensions is highly simple. It has no degrees of freedom and the constraint equations can be explicitly  solved. The main motivation to study three dimensional gravity was quantization as a toy model. For this purpose,  the Einstein-Hilbert action was written as a Chern-Simons action with a suitable gauge group \cite{wit1}, \cite{town}. Very recently \cite{car}, this property has been used to construct highly nontrivial exact solutions of the Einstein field equations with cosmological constant. The method presented here uses the zero curvature formalism of the integrable systems. The soliton connection is the Chern-Simons gauge potential with $SL(2,R)$ group as the gauge group. The soliton equations corresponding to this gauge connection are known as the AKNS equations containing the nonlinear Schrodinger and modified Korteweg de Vries equations.

\vspace{0.5cm}
\noindent
In this work we shall show that the field equations

$$
R_{\mu \nu}-\frac{1}{2} R g_{\mu \nu}=\Lambda g_{\mu \nu},
$$
where $\Lambda $ is the cosmological constant  and $\mu ,\nu=1,2, \cdots D$ are integrable when $D=3$, by using the Newmann-Penrose null tetrad formalism.
We find that SL(2,R) valued tetrad one form is the difference of two zero curvature connections without referring to the Chern-Simons theory.

\vspace{0.5cm}
\noindent
Before starting to write the gravitational field equations in null tetrad formalism we give a brief introduction to the sl(2,R) valued soliton connections and AKNS sytem. Let $x^{\mu}=(t,x,y)$ be the local coordinates of $(M,g)$  and the soliton connection ${\bf a}$ depends on the coordinates $t$ and $x$. Hence
\[
{\bf a}=P\,d x+Q\, dt,
\]
where $P(t, x)$ and $Q(t,x)$ are some sl(2,R) matrices. They are given by
\[
P=\left(\begin{array}{ll} 2 \xi & p(t,x) \cr
                         q(t,x) &-2\xi
                         \end{array} \right),
Q=\left(\begin{array}{ll} A & B \cr
                         C &-A
                         \end{array} \right)
\]
Here $\xi$ is the eigenvalue parameter and $p$ and $q$ are dynamical variables which satisfy nonlinear integrable system of equations of evolutionary type. The functions $A$,$B$, and $C$ depend on the dynamical variables $p$ and $q$ and their partial derivatives with respect to $x$.

Since the soliton connection ${\bf a}$ has no curvature then matrices $P$ and $Q$ satisfy the zero curvature condition

\[
P\,_{,t}-Q\,_{,x}+[P, Q]=0,
\]
which gives the well known AKNS equations \cite{abl}
\begin{eqnarray}
&&A_{x}-p\,C+q\, B=0, \nonumber\\
&&q_{t}+ C_{x}-2 q A-2 \xi C  =0, \nonumber\\
&&p_{t}+ B_{x}+2 p A+2 \xi B =0, \nonumber
\end{eqnarray}
We expand $A$, $B$ and $C$ as polynomials of the eigenvalue parameter $\xi$, i.e.,
\[
A=\sum_{n=0}^{N}\, A_{n}\, \xi^{N-n},~~B=\sum_{n=0}^{N}\,B_{n}\,
\xi^{N-n},~~C=\sum_{n=0}^{N}\,C_{n}\, \xi^{N-n},
\]
where the functions $A_{n}$. $B_{n}$ and $C_{n}$  depend on the dynamical variables $p$ and $q$ and their partial derivatives withe respect to $x$.

\vspace{0.5cm}
\noindent
{\bf The AKNS System}: Taking $N=2,3$ we obtain the Nonlinear Schrodinger and modified Korteweg de Vries systems respectively. This means that
when we begin with the Lax pair in $sl(2,R)$ algebra and assume them as a polynomial of the spectral parameter of degree less or equal to three then we obtain the following system of evolution equations :
\begin{eqnarray}
&& q_{t}=a_{2}\, (-\frac{1}{2}\, q_{xx}+q ^2\, p)+ a_{3}\,(-\frac{1}{4}\, q_{xxx} +\frac{3}{2}\, q p q_{x}),  \\
&& p_{t}=a_{2}\, (\frac{1}{2}\, p_{xx}-q \, p^2)+ a_{3}\,(-\frac{1}{4}\, p_{xxx}+\frac{3}{2}\, q p p_{x}).
\end{eqnarray}
Here $a_{2}$ and $a_{3}$ are arbitrary constants. For general $N$ we have a recursion relation $u_{t}={\cal R}^{N} u_{x}$ ($N=2, 3\cdots $) where $u=(p,q)^{T}$ and ${\cal R}$ is the recursion operator of the AKNS system. All the members of the AKNS system have infinitely many conserved quantities, infinitely many symmetries, bi-Hamiltonian structure, and admit soliton solution generating techniques etc.

\section{Null tetrad approach}
Newmann-Penrose formalism \cite{et} in four dimensions is much elegant when written in sl(2, C) algebra. It is known that the Einstein field equations admit prolonged Cartan frames \cite{met} in this formalism. We start first with SL(2,C) valued tetrad one form in four dimensions

\begin{equation}
\sigma=\left(\begin{array}{lll} {\bf{l}} & {\bf m} \cr
                           {\bf \bar{m}} & {\bf n} \cr
                         \end{array} \right),
\end{equation}

where ${\bf {l}}, {\bf n}, {\bf m}, \bar{\bf m} $ are null tetrad one forms. The metric takes the form
\begin{equation}
ds^2={\bf {l}} \oplus {\bf n}+{\bf n} \oplus {\bf {l}}+{\bf m} \oplus {\bf \bar{m}}+{\bf \bar{m}} \oplus {\bf m}.
\end{equation}

Here a bar over a letter denotes complex conjugation.
In our approach we will use the following SL(2,C) tetrad one form

\begin{equation}
\tilde{\sigma}=\left(\begin{array}{lll} {\bf n} & -{\bf \bar{m}} \cr
                           -{\bf m} & ~~~{\bf {l}} \cr
                         \end{array} \right)=\varepsilon \sigma \varepsilon^{-1},
\end{equation}
where

\begin{equation}
\varepsilon=\left(\begin{array}{lll} 0 & 1 \cr
                           -1 & 0 \cr
                         \end{array} \right) \label{eps}.
\end{equation}

\noindent
{\bf Spin Connection:} $w$

\begin{equation}
d {\tilde \sigma}=- w\, {\tilde \sigma}+{\tilde \sigma}\, w^{\dagger},
\end{equation}
where $ w$ is the sl(2,C) valued connection one form given as

\begin{equation}
w =\left(\begin{array}{lll} w_{0} & w_{2}\cr
                           w_{1} & -w_{0} \cr
                         \end{array} \right),
\end{equation}
where
\begin{eqnarray}
&&w_{0}=\gamma {\bf l}+\epsilon {\bf n}-\alpha {\bf m}-\beta \bar{{\bf m}}, \\
&&w_{1}=-\tau {\bf l}+-\kappa  {\bf n}+\rho  {\bf m}+\sigma  \bar{{\bf m}}, \\
&&w_{2}=\nu {\bf l}+\pi {\bf n}-\lambda {\bf m}-\mu \bar{{\bf m}},
\end{eqnarray}
where $\gamma,\epsilon, \alpha, \beta, \cdots $ are the Newman-Penrose spin coefficients. In the construction of the metric tensor we don't need these spin coefficients. In the sequel of the text we may use these Greek letters for different purposes.

\vspace{0.5cm}
\noindent
{\bf Curvature two form:} ${\bf  R}=d  w +  w \, w$

\begin{equation}
{\bf R}=\left(\begin{array}{lll} R_{0} & R_{2}\cr
                           R_{1} & -R_{0} \cr
                         \end{array} \right),
\end{equation}
where
\begin{eqnarray}
&&R_{0}=(\Lambda-\Phi_{11}+\psi_{2}) {\bf l \, n}+ \psi_{3} {\bf l m} +\Phi_{12} {\bf n \bar{m}} -\psi_{1} {\bf n \bar{m}} \nonumber \\
&&+(\psi_{2}-\Phi_{11}- \Lambda) {\bf m \bar{m}}, \\
&&R_{1}=-(\psi_{1}+ \Phi_{01}) {\bf l\, n}-(\psi_{2}+2 \Lambda) {\bf l \,m} -\Phi_{02} {\bf l \bar{m}}+\Phi_{00} {\bf n m}+\psi_{0} {\bf n \bar{m}} \nonumber \\
&&+(\Phi_{01}-\psi_{1}) {\bf m \bar{m}} \\
&&R_{2}= (\psi_{3}+\Phi_{21}) {\bf l \,n}-(\psi_{2}+2\Lambda) {\bf n \bar{m}}-\Phi_{20} {\bf n m}+ \Phi_{22} {\bf l \,\bar{m}}+\psi_{4} {\bf l m} \nonumber \\
&&-(\Phi_{21}-\psi_{3}) {\bf m \bar{m}},
\end{eqnarray}
where $\Phi_{ij}$ ($i,j=0,1,2$) are the trace free Ricci spin coefficients, $\psi_{a}$ ($a=0,1,2,3,4$) are the Weyl spinors and $\Lambda =-R/6$, where $R$ is the scalar curvature.
Let $\Phi_{ij}=0$ and $\psi_{a}=0$ for all ($i,j=0,1,2$) and  for all ($a=0,1,2,3,4$),then

\begin{equation}
{\bf R}= \Lambda \left(\begin{array}{lll} {\bf l n}-{\bf m \bar{m}}  & -2 {\bf n \bar{m}}\cr
                           -2 {\bf l m} & -{\bf l n} +{\bf m \bar{m}}\cr
                         \end{array} \right).
\end{equation}
Now let
\begin{equation}
\Gamma_{\pm}= w \pm \ell_{0} \tilde{\sigma} \varepsilon,
\end{equation}
where $\ell_{0}$ is any constant. Curvatures $\Omega_{\pm}$ of the connection $\Gamma_{\pm}$ are given by
\begin{equation}
\Omega_{\pm}={\bf R} +\ell_{0}^2 \tilde{\sigma} \varepsilon\,\tilde{\sigma} \varepsilon \pm \ell_{0} \tilde{\sigma} \left( w^{\dagger} \varepsilon+\varepsilon \, w \right).
\end{equation}
To have $\Omega_{\pm}=0$ we must have $\bar{w}=w$ and  ${\bf \bar{m}}={\bf m}$ . In this case the group SL(2,C) reduces to SL(2,R) and dimension reduces to three and
\begin{equation}
\ell_{0}^2 =-\Lambda.
\end{equation}

As a result we end up with a theorem saying that, the integrability of 4 dimensional gravity  implies that null tetrad must be real and
the spacetime dimension falls down to three and hence
SL(2,C) group is contracted to SL(2,R). This is an In{\" o}n{\" u}-Wigner contraction \cite{inonu}.

\section{Newman-Penrose Formalism in Three Dimensions}
Following \cite{anut} SL(2,R) valued tetrad one form is given by

\begin{equation}
\tilde{\sigma}=\left(\begin{array}{lll} {\bf n} & -\frac{1}{\sqrt{2}} {\bf m} \cr
                           -\frac{1}{\sqrt{2}} {\bf m} & ~~~~~{\bf {l}} \cr
                         \end{array} \right)=\varepsilon \sigma \varepsilon^{-1}
\end{equation}

\noindent
{\bf Connection one form:} $w$

\begin{equation}
d {\tilde \sigma}=-w\, {\tilde \sigma}+{\tilde \sigma}\,w^{t},
\end{equation}
where $w$ is the sl(2,R) valued connection one form given as

\begin{equation}
w=\left(\begin{array}{lll} w_{0} & w_{2}\cr
                           w_{1} & -w_{0} \cr
                         \end{array} \right),
\end{equation}
where
\begin{eqnarray}
&&w_{0}=\frac{1}{2}\,(-\epsilon^{\prime} {\bf l}+\epsilon {\bf n}-\alpha {\bf m}), \\
&&w_{1}= \frac{1}{\sqrt{2}}\,(-\tau {\bf l}-\kappa  {\bf n}+\sigma {\bf m}), \\
&&w_{2}= \frac{1}{\sqrt{2}}\,(-\kappa^{\prime} {\bf l}-\tau^{\prime} {\bf n} +\sigma^{\prime} {\bf m}),
\end{eqnarray}
where $\gamma,\epsilon, \alpha, \beta, \cdots $ are the Newman-Penrose spin coefficients. Then
\begin{eqnarray}
&&d {\bf \ell}=-\epsilon {\bf l n}+(\alpha-\tau) {\bf l m}+\kappa^{\prime} {\bf n m}, \\
&&d n=\epsilon^{\prime} {\bf l n} -\kappa^{\prime} {\bf l m}-(\alpha+\tau^{\prime}){\bf n m}, \\
&& d m=(\tau^{\prime}-\tau) {\bf l n}-\sigma^{\prime} {\bf l m}-\sigma {\bf n m}.
\end{eqnarray}

\vspace{0.5cm}
\noindent
{\bf Curvature two form:} ${\bf  R}=d w +w\, w$

\begin{equation}
{\bf R}=\left(\begin{array}{lll} R_{0} & R_{2}\cr
                           R_{1} & -R_{0} \cr
                         \end{array} \right),
\end{equation}
where
\begin{eqnarray}
&&R_{0}=(2 \Phi_{11}+\Lambda)\, {\bf l \, n}-\Phi_{12}\, {\bf l m} +\Phi_{10}\,{\bf n m}, \nonumber \\
&&R_{1}=-\sqrt{2} \Phi_{01}\,  {\bf l\, n}+\frac{1}{\sqrt{2}} (\Phi_{02}-2\,\Lambda)\,{\bf l \,m} -\sqrt{2} \Phi_{00}\,  {\bf n m}, \nonumber \\
&&R_{2}= \sqrt{2} \Phi_{12} \, {\bf l\, n}-\sqrt{2} \Phi_{22} \, {\bf l \,m} +\frac{1}{\sqrt{2}} (\Phi_{02}-2\,\Lambda)\, {\bf  n m}, \nonumber
\end{eqnarray}
where $\Phi_{ij}$ ($i,j=0,1,2$) are the trace free Ricci spin coefficients and $\Lambda =-R/6$.
Let $\Phi_{ij}=0$  for all ($i,j=0,1,2$), then

\begin{equation}
{\bf R}= \Lambda \, \left(\begin{array}{lll} {\bf l n} & -\sqrt{2}\, {\bf n m}\cr
                           -\sqrt{2}\, {\bf l m} & -{\bf l n} \cr
                         \end{array} \right),
\end{equation}
and
\begin{equation}
\tilde{\sigma}\, \varepsilon=\left(\begin{array}{lll} \frac{1}{\sqrt{2}}\,{\bf m} &~~~~ {\bf n} \cr
                           -{\bf {l}} &  -\frac{1}{\sqrt{2}}\,{\bf m}  \cr
                         \end{array} \right).
\end{equation}
Hence
\begin{equation}
\tilde{\sigma}\, \varepsilon\,\tilde{\sigma}\, \varepsilon =\left(\begin{array}{lll} {\bf l n} & -{\sqrt{2}}\,{\bf n m} \cr
                           -{\sqrt{2}}\,{\bf l m} & -{\bf {l n}} \cr
                         \end{array} \right)
\end{equation}

\vspace{0.5cm}
\noindent
Now let
\begin{equation}
\Gamma_{\pm}=w \pm \lambda \tilde{\sigma} \varepsilon,
\end{equation}
where $\lambda$ is any constant. Curvatures $\Omega_{\pm}=d \Gamma_{\pm}+\Gamma_{\pm}\, \Gamma_{\pm}$ of the connections $\Gamma_{\pm}$ are found as
\begin{equation}
\Omega_{\pm}={\bf R} +\lambda^2 \tilde{\sigma} \varepsilon\,\tilde{\sigma} \varepsilon \pm \lambda \tilde{\sigma} \left(w^{t}\, \varepsilon+\varepsilon \,w \right)
\end{equation}
Since $w$ is an sl(2,R) connection then $w^{t} \varepsilon+\varepsilon w=0$ . Taking $\lambda^2 =-\Lambda.$ then

$${\bf R} +\lambda^2 \tilde{\sigma} \varepsilon\,\tilde{\sigma} \varepsilon =0$$ Hence both $\Gamma_{+}$ and $\Gamma_{-}$ are zero curvature sl(2,R) connections.
Then three dimensional gravity is integrable and
as a result we have the soliton sl(2,R) connections

\vspace{0.5cm}
\noindent
$$\Gamma_{\pm}=w \pm \lambda \tilde{\sigma} \varepsilon$$
are the zero curvature connections for three dimensional AdS spacetimes. Writing $\Gamma_{\pm}$ in a suitable form for the integrable systems we have the tetrad one form
\begin{equation}
\tilde{\sigma}\, \varepsilon=\frac{1}{2 \lambda} (\Gamma_{+}-\Gamma_{-})\,  \label{metric}
\end{equation}

\vspace{0.5cm}
\noindent
Here in our approach the connections $\Gamma_{\pm}$ are any zero curvature soliton connections, not coming from a Chern-Simons theory. Hence they do not depend on any boundary conditions as assumed by  Cardenas et al \cite{car}.
Now we shall relate the flat gauge potential one forms $\Gamma_{\pm}$ to the well known AKNS soliton connection.
 Let $\Gamma_{+}=b_{+}^{-1}\,db_{+}+b_{+}^{-1}\, {\bf a}^{+} \, b_{+}$ where $b_{+}$ is nonsingular matrix and ${\bf a}^{+}$ is a soliton connection one form. Then $\Omega_{\pm}=d\Gamma_{\pm}+\Gamma_{\pm} \, \Gamma_{\pm}=0$
 implies that ${\bf a}^{+}$ is also a flat connection
 \[
 d{\bf a}^{+}+{\bf a}^{+}\, {\bf a}^{+}=0
 \]
 Similarly let $\Gamma_{-}= \,b_{-}^{-1}\, d b_{-}+b^{-1}_{-}\, {\bf a}^{-} \, b_{-}$ where $b_{-}$ is another nonsingular matrix and
 \[
 d {\bf a}^{-}+{\bf a}^{-}\, {\bf a}^{-}=0.
 \]
 Choosing
\begin{equation}
\Gamma_{\pm} =\left(\begin{array}{lll} X_{\pm}& Y_{\pm} \cr
                           Z_{\pm} & -X_{\pm} \cr
                         \end{array} \right)
\end{equation}
where $X_{\pm}$, $Y_{\pm}$ and $Z_{\pm}$ are one forms. Then from (\ref{metric}) we find that
\begin{eqnarray}
&&{\bf m}=\frac{1}{\sqrt{2} \lambda} \left(X_{+}-X_{-}\right), \label{tet1}\\
&&{\bf n}=\frac{1}{2 \lambda} \left(Y_{+}-Y_{-}\right), \label{tet2}\\
&& {\bf l}=-\frac{1}{2 \lambda} \left(Z_{+}-Z_{-}\right), \label{tet3}
\end{eqnarray}

\section{Soliton Connection and the Metric}
Let $x^{\mu}=(t,x,y)$ be the local coordinates of $(M,g)$ and the matrices $b_{\pm}$ depend  on all coordinates and the soliton connections $a^{\pm}$ depend on the coordinates $t$ and $x$. Hence
\[
{\bf a}^{\pm}=P^{\pm}\,d x+Q^{\pm}\, dt
\]
where $P(t, x)$ and $Q(t,x)$ are some $sL(2,R)$ matrices. They are given by
\[
P^{\pm}=\left(\begin{array}{ll} 2 \xi^{\pm} & p^{\pm}(t,x) \cr
                         q^{\pm}(t,x) &-2\xi^{\pm}
                         \end{array} \right),
Q^{\pm}=\left(\begin{array}{ll} A^{\pm} & B^{\pm} \cr
                         C^{\pm} &-A^{\pm}
                         \end{array} \right).
\]
Here $\xi^{\pm}$ are the eigenvalue parameters and $p^{\pm}$ and $q^{\pm}$ are dynamical variables which satisfy nonlinear integrable system of equations of evolutionary type. The functions $A^{\pm}$, $B^{\pm}$ and $C^{\pm}$  depend on the dynamical variables $p^{\pm}$ and $q^{\pm}$ and their partial derivatives withe respect to $x$.
Choosing the b matrices as

\begin{equation}
b =\left(\begin{array}{lll} \alpha & \beta \cr
                           \gamma & \sigma \cr
                         \end{array} \right),
\end{equation}
with unit determinant $\alpha \sigma-\beta \gamma=1$. Functions $\alpha$, $\beta$, $\gamma$ and $\sigma$ depend on all coordinates $t,x,y$.
For each zero curvature connection $\Gamma_{+}$ and $\Gamma_{-}$ we use different gauge matrices as $b_{\pm}$. Then we find the one forms $X$, $Y$ and $Z$ (ignoring $\pm$ subscripts) as

\begin{eqnarray}
&& X= \left[2(\alpha \sigma +\beta \gamma)\xi+\sigma \gamma p -\alpha \beta q \right] dx+\left[(\alpha \sigma+\beta \gamma) A+\sigma \gamma B-\alpha \beta C \right]\,dt+\sigma d \alpha-\beta  d \gamma, \nonumber\\
&&Y=\left[4 \sigma \beta \xi+\sigma^2 p-\beta^2 q \right ] dx+ \left[2 \sigma \beta A+\sigma^2 B-\beta^2 C \right] dt+\sigma d \beta-\beta d \sigma,\nonumber\\
&&Z=\left[-4 \xi \gamma \alpha -\gamma^2 p+\alpha^2 q \right] dx+\left[-2 \alpha \gamma A-\gamma^2 B+\alpha^2 C \right] dt-\gamma  d \alpha+\alpha d \gamma. \nonumber
\end{eqnarray}
Finally by using the above expressions we obtain the tetrad one forms ${\bf l}$, ${\bf n}$, ${\bf m}$ from (\ref{tet1})-(\ref{tet3}) and the metric tensor is given by

\begin{equation}
g_{\mu \nu}=l_{\mu}\, n_{\nu}+l_{\nu}\,n_{\mu}-m_{\mu}\, m_{\nu}.
\end{equation}

\vspace{0.5cm}
\noindent
The only gravitational field equations are the following six AKNS equations

\begin{eqnarray}
&&A^{\pm}_{x}-p^{\pm}\,C^{\pm}+q^{\pm}\, B^{\pm}=0, \\
&&q^{\pm}_{t}+ C^{\pm}_{x}-2 q^{\pm} A^{\pm}-2 \xi^{\pm} C^{\pm}  =0, \\
&&p^{\pm}_{t}+ B^{\pm}_{x}+2 p^{\pm} A^{\pm}+2 \xi^{\pm} B^{\pm} =0.
\end{eqnarray}

\section{Connections with Non-vanishing Curvatures}

Let $\Gamma_{+}$ and $\Gamma_{-}$ be two sl(2,R) valued connection one forms with curvatures $\Omega_{+}$ and $\Omega_{-}$ respectively. The three dimensional spacetime geometry has an SL(2,R) valued  null tetrad one form $\tilde{\sigma}$ and sl(2,R) valued connection one form $w$  given in terms of the connections $\Gamma_{+}$ and $\Gamma_{-}$ as follows

\begin{equation}
\tilde{\sigma}=\frac{1}{2 \lambda} (\Gamma_{+}-\Gamma_{-})\, \varepsilon,~~~w=\frac{1}{2}\,(\Gamma_{+}+\Gamma_{-}), \label{trans}
\end{equation}
where the matrix $\varepsilon$ is defined in (\ref{eps}) and $\lambda $ is a constant. Then we find that
\begin{equation}
d \tilde{\sigma}+w \tilde{\sigma}-\tilde{\sigma} w^{t}=T, \label{denk1}
\end{equation}
where $T$ is the sl(2,R) valued torsion two form and found as $T=\Omega_{+}-\Omega_{-}$. The curvature two form ${\bf R}$ of the spacetime geometry is found as
\begin{equation}
{\bf R}+\lambda^2\, \tilde{\sigma} \varepsilon \tilde{\sigma} \varepsilon=\frac{1}{2}(\Omega_{+}+\Omega_{-}). \label{denk2}
\end{equation}
The last two equations (\ref{denk1}) and (\ref{denk2}) imlpy a 3 dimensional gravity with torsion and nonzero matter. When we let torsion to vanish then the connections $\Gamma_{+}$ and $\Gamma_{-}$ have the same curvature, i.e., $\Omega_{+}=\Omega_{-}$. Let us call this curvature as $\Omega$. Hence the associated three dimensional gravity field equations are
\begin{eqnarray}
&&d \tilde{\sigma}+w \tilde{\sigma}-\tilde{\sigma} w^{t}=0, \\
&&{\bf R}+\lambda^2\, \tilde{\sigma} \varepsilon \tilde{\sigma} \varepsilon=\Omega.
\end{eqnarray}
By choosing connections $\Gamma_{+}$ and $\Gamma_{-}$ properly the above field equations may correspond to a well defined non-vacuum case. We may consider the equations (\ref{trans}) as Backlund transformations from vacuum to non-vacuum solutions of the Einstein field equations.

\section{Conclusion}
Using the null tetrad formalism in three dimensions we showed the the vacuum equations with cosmological constant are integrable.
This means that we determine the tetrad one forms in terms of the curvature free connection one forms. Choosing the zero curvature connection one forms (soliton connections) properly we find the metric of the spacetime in terms of the variables of the well known AKNS system including the nonlinear Schrodinger, KdV and Modified KdV equations. In our formalism the gauge matrices $b_{\pm}$ are taken independent and  depend on all coordinates $t,x,y$. In \cite{car} these matrices are assumed to be the inverses of each other and depend only on one coordinate. Some of these functions can be eliminated by using the SL(2,R) gauge transformation but we keep them for having the metric with maximum number of free functions. Since we have derived the tetrad one form $\tilde{\sigma}$ in terms of the zero curvature connections $\Gamma_{\pm}$ without referring to the Chern-Simons theory we do not necessarily need any boundary conditions to be satisfied by the tetrad functions.

We have some number of comments and questions to be answered. One of them is  the infinite number of symmetries and conservation laws of the AKNS system.
How these be perceived in the three dimensional gravity is unclear. Another point is the recursion operator of the AKNS system. By  using the recursion operator of AKNS system one can go from one system to another one. How the corresponding three dimensional metrics are related is another puzzle. Finally, how we should interpret  solitons and soliton collisions in three dimensional gravity needs further efforts.

In the last part of this paper we have considered the connections with non-vanishing curvatures and discussed possibility of obtaining non-vacuum solutions.

\section{Acknowledgment}

I would like to thank Bayram Tekin and Tahsin Ça\u{g}r\i{} \c{S}i\c{s}man for their kind interest in this work.


\section*{References}
\begin{thebibliography}{9}
\bibitem{wit1} Witten E, {\it $2 + 1$ dimensional gravity as an exactly soluble system}, Nucl. Phys. {\bf B 311},46-78
(1988)

\bibitem{town} Achocarro A and Townsend P.K,{\it A Chern-Simons Action For Three-Dimensional Anti-de Sitter Supergravity Theories}, Phyics Letters {\bf 180B} (1986).


\bibitem{car}  Cardenas M, Correa F, Lara K, and Pino1 M, {\it Integrable Systems and Spacetime Dynamic}, Physical Review Letters {\bf 127}, 161601 (2021).
( arXiv:2104.09676v2)


\bibitem{abl} Ablowitz M.J, Kaup D.J, Newell A.C, and Segur H, {\it  Nonlinear-Evolution Equations of Physical Significance},
Phys. Rev. Lett. 31, 125 (1973).

\bibitem{et} Newman E.T and Penrose R, {\it An approach to gravitational radiation by a method of spin coefficients }, Journ. Math. Phys. {\bf 3}, 566  (1962); ibid {\bf 4}, 998 (1963).

\bibitem{met} G{\" u}rses M, {\it Prolongation Structure and a Backlund Transformation for Vacuum Field Equations}, Physics Letters {\bf 101A} , (1984) 388.

\bibitem{inonu} In{\" o}n{\" u} E, Wigner E.P, {\it on the Contraction of Groups and Their Representations}. Proc. Natl. Acad. Sci. {\bf 39}, 510-24 (1953).

\bibitem{anut}  Aliev A.N and Nutku Y, {\it Spinor Formulation of Topologically Massive Gravity}, Class. Quantum Gravity {\bf 12},2914-2925 (1995)
\end{thebibliography}
\end{document}